\documentclass[aps,prc,preprint,superscriptaddress]{revtex4-1}

\usepackage{graphicx}
\usepackage{amsmath}
\usepackage{multirow}
\usepackage{tikz}
\usepackage{bm}
\usepackage{color}
\usepackage{hyperref}
\begin{document}


\title{Chiral NNLO$_{\text{sat}}$ descriptions of nuclear multipole resonances
within the random phase approximation}


\author{Q. Wu}
\author{B. S. Hu}
\author{F. R. Xu}
\email{frxu@pku.edu.cn}
\author{Y. Z. Ma}
\author{S. J. Dai}
\affiliation{School of Physics, and State Key Laboratory of Nuclear Physics and Technology, Peking University, Beijing 100871, China}
\author{Z.H. Sun}
\affiliation{School of Physics, and State Key Laboratory of Nuclear Physics and Technology, Peking University, Beijing 100871, China}
\affiliation{Department of Physics and Astronomy, University of Tennessee, Knoxville, Tennessee 37996, USA}
\affiliation{Physics Division, Oak Ridge National Laboratory, Oak Ridge, Tennessee 37831, USA}
\author{G. R. Jansen}
\thanks{This manuscript has been authored by UT-Battelle, LLC under
  Contract No. DE-AC05-00OR22725 with the U.S. Department of
  Energy. The United States Government retains and the publisher, by
  accepting the article for publication, acknowledges that the United
  States Government retains a non-exclusive, paid-up, irrevocable,
  world-wide license to publish or reproduce the published form of
  this manuscript, or allow others to do so, for United States
  Government purposes. The Department of Energy will provide public
  access to these results of federally sponsored research in
  accordance with the DOE Public Access
  Plan. (http://energy.gov/downloads/doe-public-access-plan).}

\affiliation{National Center for Computational Sciences, Oak Ridge National
Laboratory, Oak Ridge, TN 37831, USA}
\affiliation{Physics Division, Oak Ridge National
Laboratory, Oak Ridge, Tennessee 37831, USA}


\date{\today}

\begin{abstract}
  We study nuclear multipole resonances in the framework of the random phase
  approximation using the chiral potential NNLO$_{\text{sat}}$. This potential
  includes two- and three-body terms that has been simultaneously optimized to
  low-energy nucleon-nucleon scattering data and selected nuclear structure
  data. Our main foci have been the isoscalar monopole, isovector dipole, and
  isoscalar quadrupole resonances of the closed-shell nuclei, $^4$He,
  $^{16,22,24}$O, and $^{40,48}$Ca. These resonance modes have been widely
  observed in experiment. In addition, we use a renormalized chiral potential
  $V_{\text{low-}k}$, based on the N$^3$LO two-body potential by Entem and
  Machleidt [Phys. Rev. \textbf{C68}, 041001 (2011)]. This introduces a dependency on the cutoff parameter used in the
  normalization procedure as reported in previous works by other groups. While
  NNLO$_{\text{sat}}$ can reasonably reproduce observed multipole resonances, it
  is not possible to find a single cutoff parameter for the $V_{\text{low-}k}$
  potential that simultaneously describe the different types of resonance modes.
  The sensitivity to the cutoff parameter can be explained by missing induced
  three-body forces in the calculations. Our results for
  neutron-rich $^{22,24}$O show a mixing nature of isoscalar and isovector
  resonances in the dipole channel at low energies. We predict that $^{22}$O
  and $^{24}$O have low-energy isoscalar quadrupole resonances at energies lower than 5 MeV.
\end{abstract}


\maketitle

\section{Introduction}
The occurrence of collective resonances is a common phenomenon of many-body
quantum systems. At excitation energies above particle thresholds, the response
of the nucleus to an external nuclear or electromagnetic field is dominated by
collective vibrations of various multipolarities. The isovector giant dipole
resonance (IVGDR) \cite{Bothe1937,PhysRev.71.3,Berman1975} was one of the
earliest collective vibrational modes observed in nuclei. In IVGDR, the protons
and neutrons oscillate against each other in a dipole mode \cite{Migdal1944331}.
Giant resonances can occur on the whole nuclear chart and are directly connected
to bulk properties of nuclei, such as incompressibility and symmetry energy
\cite{Grag_Acta,PATEL2012447,GUPTA2016482,refId0}. For neutron-rich isotopes,
there exist electric dipole ($E1$) responses at low energy with weak strengths,
named pigmy dipole resonances (PDR) \cite{doi:10.1139/p69-348}. These are
interpreted as dipole oscillations of excess neutrons against the $N=Z$
proton-neutron saturated core \cite{PhysRevC.3.1740}. Furthermore, in weakly-bound
nuclei along the driplines, it has been speculated that there exist resonances
called soft dipole resonances \cite{0295-5075-4-4-005}, a dipole mode where loosely-bound
nucleons oscillate against a core. The soft resonances can occur at even lower
energies, attracting much interest in both theory
\cite{SDR,PhysRevC.89.064303,0034-4885-70-5-R02} and experiments
\cite{Savran2013210,PhysRevLett.114.192502,Tanihata2013215,Savran2013210}. A soft
dipole resonance has been observed in the halo nucleus $^{11}$Li at an
excitation energy of only 1.03 MeV \cite{PhysRevLett.114.192502}. Low-energy
monopole and quadrupole resonances have also been observed in neutron-rich
nuclei~(see e.g. Refs.
\cite{PhysRevLett.113.032504,PhysRevC.92.014330,SPIEKER2016102}). A microscopic
description of multipole resonances, based on realistic nuclear forces, is
challenging because of the computational complexity involved in solving the
nuclear many-body problem.

Within the Hartree-Fock (HF) approach with non-relativistic Skyrme or
relativistic meson-exchange potential, the random phase approximation (RPA)
which describes nuclear collective vibrations by particle-hole excitations, has
been successfully applied to describe multipole responses in nuclei (see e.g.
the reviews \cite{0954-3899-38-3-033101,pts016,0034-4885-70-5-R02} and
references therein). However, details of the calculations depend on nuclear
forces used. Especially, the calculations of low-lying resonance strengths can
be drastically different with different forces. For example, it was pointed out
that the tensor force in the Skyrme interaction has a significant effect on charge-exchange dipole excitations \cite{PhysRevLett.105.072501}, while the tensor force is missing in many other calculations. In theory, the collective responses of nuclei are directly related to certain properties of the underlying nuclear force. 

Recent nuclear structure calculations have highlighted the use of realistic
nuclear forces that accurately describe nucleon-nucleon scattering. The giant
dipole resonances (GDR) of $^3$H, $^3$He, and $^4$He, have been
investigated within the framework of the correlated hyperspherical harmonic
expansion \cite{EFROS2000223, PhysRevLett.96.112301}, using the Argonne
AV18~\cite{PhysRevC.51.38} two-body
(NN) potential plus the Urbana~\cite{PhysRevC.56.1720} three-body (NNN) force. It was found that
three-body forces had a strong effect in the region of excitation energies
higher than 50 MeV.
The $^4$He giant resonances were further studied by the no core shell model
\cite{QUAGLIONI2007370} and with the effective interaction hyperspherical
harmonics  method \cite{PhysRevC.91.024303}, using interactions from chiral effective field theory
(EFT).
In heavier nuclei, the random phase approximation based on realistic forces has been successful.
In Refs.
\cite{Paar2006,Papakonstantinou2007,Papakonstantinou2010,PhysRevC.83.064317},
a correlated interaction derived from the AV18 potential by using the unitary
correlation operator method (UCOM)~\cite{ROTH20043} was used with HF-RPA. These calculations 
reproduced experimental multipole resonances reasonably well, but the details
depended on the UCOM cutoff parameter of the potential.

The dipole resonances of closed-shell nuclei have also been studied using the
coupled-cluster approach with chiral forces
\cite{PhysRevLett.111.122502,PhysRevC.90.064619}. Chiral effective field theory
offers a consistent framework to derive two- and three-nucleon forces. Recently,
the Oak Ridge group suggested a chiral EFT potential called NNLO$_{\text{sat}}$ where
two-nucleon and three-nucleon forces were optimized simultaneously to low-energy
nucleon-nucleon scattering data, as well as binding energies and radii of
few-nucleon systems and selected isotopes of carbon and oxygen
\cite{PhysRevC.91.051301}. This potential has been successfully used to describe the electric dipole polarizabilities of $^4$He, $^{40}$Ca and $^{16, 22}$O \cite{Hagen:2016aa,PhysRevC.94.034317}. 

The RPA has been proved to be an efficient approximation to describe collective
resonances and long-range correlations. The low computational cost makes it
feasible to describe multipole resonances of medium-mass nuclei. These resonances are considered using a
one-body operator response of the nucleus, thus the resonance strength should
come mainly from one-particle one-hole ($1p1h$) excitations. In this paper, we perform
self-consistent HF-RPA calculations with the chiral NNLO$_{\text{sat}}$
potential and describe multipole resonances of closed-shell helium, oxygen and
calcium isotopes. In this context, self-consistence means that the HF and RPA calculations
are performed using the same interaction. However, the three-nucleon force has
only been included fully in the Hartree-Fock calculations, while in the RPA
calculations we keep only the normal-ordered zero-, one-, and two-body terms of
the three-body force~\cite{PhysRevLett.109.052501,PhysRevC.76.034302}. For comparison, we also performed HF-RPA calculations
with a chiral two-body N$^3$LO potential \cite{Entem2003,Machleidt2011a}
renormalized by the $V_{\text{low-}k}$ technique \cite{Bogner2003} at different
low-momentum cutoff parameters. 

\section{Theoretical framework}
\subsection{The RPA formulation}
The RPA equations can be derived within different theoretical frameworks such as Green's
function theory, the small amplitude limit of time-dependent Hartree Fock (TDHF) and
equation of motion (EOM). In the framework of EOM, the details of the RPA
formulation can be found in a standard textbook \cite{nuclear_ring}. In this paper, we briefly
state the RPA equations in an angular momentum coupled representation. The $A$-body intrinsic Hamiltonian can be written as
\begin{eqnarray}
  H&=&\sum\limits_{i=1}^{A}{\frac{{\bm p}_i^2}{2m}} + \sum\limits_{i<j=1}^A V_{ij} -
  \frac{(\sum\limits_{i=1}^{A}{{\bm p}_i})^2}{2mA} + \sum\limits_{i<j<k=1}^A W_{ijk} \nonumber\\
  &=&(1-\frac{1}{A})\sum\limits_{i=1}^{A}{\frac{{\bm p}_i^2}{2m}} +
  \sum_{i<j=1}^A(V_{ij}-\frac{{\bm p}_i\cdot {\bm p}_j}{mA}) + \sum\limits_{i<j<k=1}^A
  W_{ijk},
  \label{eq:H}
\end{eqnarray}
where $m$ is the average mass of a nucleon, $A$ is the mass number of the
nucleus, ${\bm p}_i$ is the nucleon momentum in the laboratory frame, $V_{ij}$ is the
two-body nucleon-nucleon interaction, and $W_{ijk}$ is the three-nucleon
interaction. In
the harmonic oscillator (HO) basis, we first perform spherical HF
calculations \cite{Hu2016} where the total angular momentum is preserved.
In this paper, we are only interested in spherically closed-shell nuclei.

We write the Hamiltonian (\ref{eq:H}) in the second-quantized formulation using the HF basis
\begin{eqnarray}
  H=\mathrm{E_{HF}}+\sum\limits_i \epsilon_i \{\hat{a}_i^\dagger \hat{a}_i\} + \frac{1}{4} \sum\limits_{ijkl} \langle ij|V|kl \rangle \{\hat{a}_i^\dagger \hat{a}_j^\dagger \hat{a}_l \hat{a}_k\},
\end{eqnarray}  
where $\mathrm{E_{HF}}$ is the HF ground-state energy, and $V$ is the
nucleon-nucleon interaction with a correction from the normal-ordered
three-nucleon interaction, $\epsilon_i$ are the HF
single-particle energies, while $\hat{a}_i^\dagger$ and $\hat{a}_i$ are
operators that respectively create and annihilate a particle in the state labelled $i$. The symbol $\{\ldots\}$ indicates the normal-ordered
form of the operators with respect to the HF ground state. 
The excited states can be written as
\begin{eqnarray}
  | \nu,JM \rangle = Q_{\nu,JM}^\dagger |0\rangle,
\end{eqnarray}
where $|0\rangle$ is the ground state. The operator $Q_{\nu,JM}^\dagger$ creates
excited states with angular momentum $J$ and its projection $M$ and other quantum
numbers $(\nu)$. In the RPA, this operator is approximated as
\begin{eqnarray}
  Q_{\nu,JM}^\dagger = \sum_{ph} \left[ X_{ph}^{\nu,JM} {A_{ph}^{JM}}^\dagger - (-1)^{J+M} Y_{ph}^{\nu,JM} A_{ph}^{J,-M}\right],
\end{eqnarray}  
where $X_{ph}^{\nu,JM}$ and $Y_{ph}^{\nu,JM}$ are forward and backward-going
particle-hole amplitudes of the state ($\nu, JM$), respectively. The labels $ph$
represents $1p1h$ excitations, and the operator ${A_{ph}^{JM}}^\dagger$ couples the $1p1h$ configuration to the angular momentum $J$ and projection $M$. The summation runs over all allowed $1p1h$ excitations with $(JM)$ in the model space. The operator ${A_{ph}^{JM}}^\dagger$ can be expressed in the HF basis as
\begin{eqnarray}
  {A_{ph}^{JM}}^\dagger=\sum_{m_p m_h} (-1)^{j_h-m_h} \langle j_p m_p,j_h~-m_h| JM \rangle \hat{a}_{j_p m_p}^\dagger \hat{a}_{j_h m_h},
\end{eqnarray}
where $j$ and $m$ are the total angular momentum and its projection of a HF single-particle state, with $p$ for a particle state and $h$ for a hole state.

The EOM \cite{Rowe1968} gives a set of coupled equations for the amplitude
vectors ${\bm X}^{\nu,J}$ and ${\bm Y}^{\nu,J}$,
\begin{eqnarray}
\label{eq:rpa_eq}
 \begin{pmatrix} \mathbf{A}^{J} & \mathbf{B}^{J}\\ - {\mathbf{B}^{J}}^* & -{\mathbf{A}^{J}}^* \end{pmatrix} \begin{pmatrix}{\bm X}^{\nu,J} \\{\bm Y}^{\nu,J}\end{pmatrix}=\hbar\Omega_\nu \begin{pmatrix} \mathbf{G}^J & \mathbf{0}\\\mathbf{0} & {\mathbf{G}^J}^* \end{pmatrix} \begin{pmatrix}{\bm X}^{\nu,J} \\{\bm Y}^{\nu,J}\end{pmatrix},
\end{eqnarray}
where $\hbar\Omega_\nu$ is the excitation energy of the $\nu$-th eigen state
$({\bm X}^{\nu,J}, {\bm Y}^{\nu,J})$. The matrices are given by their matrix
elements
\begin{eqnarray}
\label{eq:rpa_gme}
\mathbf{A}^{J}_{ph,p'h'}&=&\langle 0| [ A^{JM}_{ph}, H, {A^{JM}_{p'h'}}^\dagger ] | 0\rangle\nonumber,\\
\mathbf{B}^{J}_{ph,p'h'}&=&-\langle 0| [A^{JM}_{ph}, H, (-1)^{J+M}A^{J,-M}_{p'h'}] | 0\rangle,\\
\mathbf{G}^{J}_{ph,p'h'}&=&\langle 0| [A^{JM}_{ph},{A^{JM}_{p'h'}}^\dagger] | 0 \rangle\nonumber,
\end{eqnarray}
where the double commutator $[A,B,C]$ is defined as
$2[A,B,C]=[A,[B,C]]+[[A,B],C]$ \cite{Rowe1968}, and $|0\rangle$ is the
correlated ground state.

The ground state in Eq.
(\ref{eq:rpa_gme}) can be approximated in different ways.
If we approximate it by the HF ground state $|0\rangle = |\text{HF}\rangle$, it
is usually known as the quasi-boson (QB) approximation, and we obtain the matrix elements as done in the standard RPA,
\begin{eqnarray}
  \label{eq:rpa_me}
  \mathbf{A}^{J}_{ph,p'h'}&=&(\epsilon_p-\epsilon_h)\delta_{pp'}\delta_{hh'} + \langle ph^{-1}, J| V |p' h'^{-1},J\rangle\nonumber,\\
  \mathbf{B}^{J}_{ph,p'h'}&=&\langle ph^{-1},J|V |p'^{-1}h',J\rangle,\\
  \mathbf{G}^{J}_{ph,p'h'}&=&\delta_{pp'}\delta_{hh'}\nonumber,
\end{eqnarray}
where $\epsilon_p$ and $\epsilon_h$ are the energies of the HF single-particle and -hole states, respectively.
The cross-coupled matrix elements are defined as follows,
\begin{eqnarray}
  \langle ac^{-1},J | V| b^{-1}d,J \rangle 
  &\equiv&(-1)^{j_b-j_d+J} \langle ac^{-1},J | V| db^{-1},J \rangle\nonumber\\
  &\equiv& \sum\limits_{J'}(2J'+1)(-1)^{j_b+j_c+J+J'}\begin{Bmatrix} j_a & j_c & J\\j_d & j_b & J'\end{Bmatrix}_{6j}\times\nonumber\\
    && \quad \sqrt{(1+\delta_{ab})(1+\delta_{cd})} \langle ab,J'|V|cd,J' \rangle,
\end{eqnarray}
where $a,b,c$ and $d$ indicate the HF single-particle states, and $\langle ab, J'|V|cd,J'\rangle$ is the coupled nucleon-nucleon interaction matrix element defined by
\begin{eqnarray}
  \langle ab,J'|V|cd,J' \rangle=\sum\limits_{m_a m_b m_c m_d} && \langle j_a m_a,j_b m_b|J' M \rangle \langle j_c m_c,j_d m_d|J' M \rangle \times\nonumber\\
   &&\frac{1}{\sqrt{(1+\delta_{ab})(1+\delta_{cd})}} \langle ab|V|cd \rangle 
\end{eqnarray}

In the QB approximation, the double commutators are calculated using the HF
ground state rather than the actual yet unknown ground state, which is somewhat
inconsistent. Efforts have been made to mitigate the problem by taking the
dispersion of the Fermi surface in the ground state, known as the extended RPA
(ERPA) \cite{Papakonstantinou2007,Gambacurta2008}. The ERPA adopts an iteration
procedure. Starting from the HF ground state, the initial amplitudes $\bm X$ and
$\bm Y$ are determined using Eqs. (\ref{eq:rpa_eq}) and (\ref{eq:rpa_me}), and
are used to calculate one-body densities of the correlated ground state \cite{Rowe1968a}. After this step, new matrix elements in Eq. (\ref{eq:rpa_gme})
which depend on the density are calculated, and then new amplitudes $\bm X$ and
$\bm Y$ are obtained. The iterative procedure is repeated until convergence is reached.
However, it turns out that the effect of the ground-state correlation on
collective multipole resonances is rather small \cite{Papakonstantinou2007}. As
commented in Ref. \cite{Rowe1968}, the calculations of the double commutators
should be less sensitive to the choice of the ground state $|0\rangle$. In this
paper we will not use this approach, but we will calculate a correction to the
ground state energy as described below.

\subsection{The ground-state correlation}
The RPA calculates relative quantities, such as excitation energy and transition
densities. It does not pay much attention to the ground state explicitly.
Following the QB approximation, the total binding energy of the ground state can
be evaluated by considering contributions from the zero-point energies of all
possible vibration modes \cite{nuclear_ring},
\begin{eqnarray}
  \label{eq:gs_energy}
  \mathrm{E_{\text{RPA-QB}}}=\mathrm{E_{HF}}-\frac{1}{2} \mathrm{Tr}~\mathbf{A} +\frac{1}{2}\sum\limits_\nu\hbar\Omega_\nu,
\end{eqnarray}
where $\mathrm{Tr}~\mathbf{A}$ is the trace of all the $\mathbf{A}^{J}$ matrices appearing in Eq. (\ref{eq:rpa_eq})-(\ref{eq:rpa_me}), and the summation is over all possible vibration modes including charge exchange excitations where isospins are changed.

\begin{figure}[!htbp]
\centering
\includegraphics[width=\columnwidth]{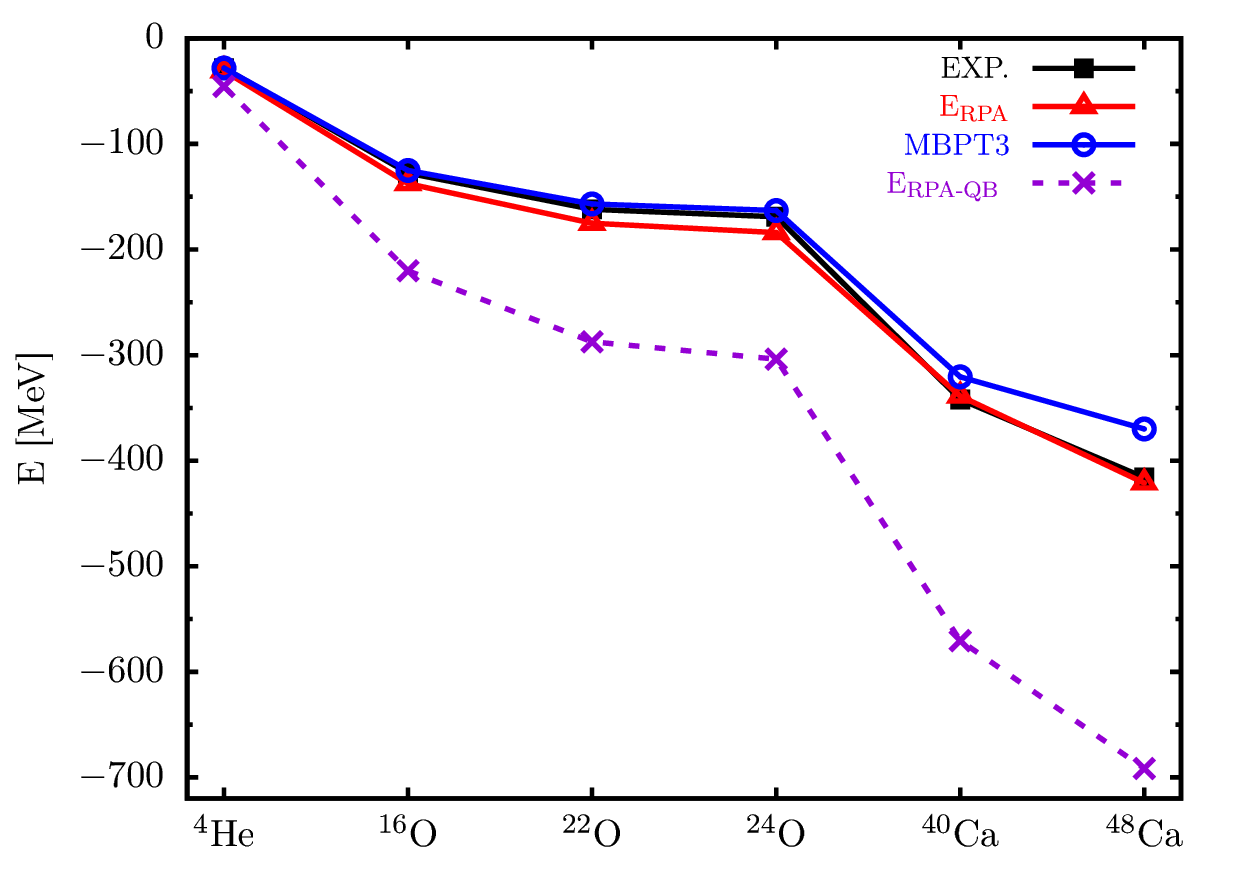}
\caption{\label{fig:binding}Calculated RPA ground-state energy with the second-order MBPT correction ($\mathrm{E_{RPA}}$),
  compared with MBPT3 calculations and experimental data. The quasi-boson RPA
  energy ($\mathrm{E_{\text{RPA-QB}}}$) given by Eq. (\ref{eq:gs_energy}) is also displayed for comparison.
The chiral NNLO$_{\text{sat}}$ interaction is used. We take $\hbar\omega=22$ MeV
for the harmonic oscillator basis and truncate the basis with $N_{\text
{max}}=2n+l=12$.}
\end{figure}

It has been known that the QB approximation overestimates the correlation
energy. In Ref. \cite{rpa_rccd} it was shown that the coupled-cluster
doubles equation including only ring terms (ring-CCD) is equivalent to RPA,
except for a factor of $1/2$ in the ring-CCD correlation energy. More precisely,
by comparing the contributions from the ring
diagrams of many body perturbation theory (MBPT) to all orders, the RPA
correlation energy from the QB approximation is overestimated by a factor of two in
the second-order perturbation diagram \cite{ELLIS1970625}. This means that the
a better approximation to the ground-state energy can be obtained by subtracting the second-order
correlation term, 
\begin{eqnarray}
  \label{eq:ring_diagram}
  \mathrm{E_{RPA}=E_{\text{RPA-QB}}-E^{(2)}},
\end{eqnarray}
where E$^{\text{(2)}}$ is the second-order perturbation diagram \cite{Hu2016}.
The binding energy calculation now includes contributions from all ring
diagrams. In energy calculations based on realistic nuclear forces, this
second-order correction plays an important role \cite{Hu2016}. Fig.
\ref{fig:binding} shows the uncorrected (E$_{\text {RPA-QB}}$) and corrected (E$_{\text {RPA}}$) binding energies, and
MBPT calculations up to third-order (MBPT3), compared with experimental data. The chiral
interaction, NNLO$_{\text{sat}}$ \cite{PhysRevC.91.051301}, was used for these
calculations, where the three-body force was truncated at the normal-ordered
two-body level in the Hartree-Fock basis. We see that the usual RPA energy given
by Eq. (\ref{eq:gs_energy}) without the $\mathrm{E^{(2)}}$ correction gives poor
results, and that the $\mathrm{E^{(2)}}$ correction plays a crucial role.

\subsection{Responses and transitions}
The nuclear response to the electromagnetic field can be described by multipole
operators ($O_{\lambda \mu}$) that come from a multipole expansion of the nuclear coupling with the
field \cite{nuclear_models}. In the long-wavelength limit, the lowest-order
electric multipole transition operators of interest are given below.

The isoscalar monopole operator is defined as
\begin{eqnarray}
O_{00}^{\text{IS}}=\sum\limits_{i=1}^{A} r_i^2,
\end{eqnarray}
where $r_i=|\bm{r}_i|$ is the distance in radius for the $i$-th nucleon.
The effective isovector dipole operator with the center-of-mass motion removed is
\begin{eqnarray}
O_{1\mu}^{\text{IV}}&=&e\frac{N}{A}\sum\limits_{i=1}^{Z} r_i Y_{1\mu}(\hat{\bm{r}}_i)-e\frac{Z}{A}\sum\limits_{i=1}^N r_i Y_{1\mu}(\hat{\bm{r}}_i)\nonumber\\
&=&e\sum\limits_{i=1}^A (\frac{N-Z}{2A}-t_z^{(i)}) r_i Y_{1\mu}(\hat{\bm{r}}_i),
\end{eqnarray}
where $N$ and $Z$ are the neutron and proton numbers, respectively, $e$ is the
charge of a proton, $Y_{\lambda\mu}(\hat{\bm{r}}_i)$ is the spherical harmonics
for the multipole mode ($\lambda\mu$), while $t_z^{(i)}$ is the isospin projection of the $i$-th nucleon.
The isoscalar dipole operator \cite{VANGIAI19811} that corresponds to the compressional dipole mode is
\begin{eqnarray}
O_{1\mu}^{\text{IS}}&=&\sum\limits_{i=1}^{A} r_i^3 Y_{1\mu}(\hat{\bm{r}}_i).
\end{eqnarray}
For the multipolarity $l\ge 2$ isoscalar operators, the usual forms are taken as
\begin{eqnarray}
O_{l\mu}^{\text{IS}}=\sum\limits_{i=1}^{A} r_i^l Y_{l\mu}(\hat{\bm{r}}_i).
\end{eqnarray}

We have omitted the charge $e$ in isoscalar operators, since the transition is
not necessarily induced by the electromagnetic interaction. In the $J^\pi=1^-$ dipole
channel, a spurious state associated with the center-of-mass motion emerges as
the HF breaks the transitional symmetry. In principle, the energy of the
spurious state is exactly zero in the RPA \cite{nuclear_ring}, which gives an
easy way to identify the spurious state. In practice, the energy of the
spurious state is not exactly zero due to a limited model space. It is, however, close to zero and well separated from other states. For example,
in our calculations with a model space of $N_{\text {max}}=12$ the energies
of the spurious states in $^4$He and $^{16}$O are as small as $10^{-15}$ MeV.
The maximum energy of the spurious state happens in $^{48}$Ca, where it is 0.8 MeV,
well separated from the lowest resonance peak at $\approx 11$ MeV. Therefore, we
can easily identify and remove the $1^-$ spurious state. 

The reduced $l$-pole electric transition probability is given by
\begin{eqnarray}
B(El,0\to\nu)&=&\sum\limits_\mu|\langle \nu | O_{l\mu} |0 \rangle |^2\nonumber\\
&\approx&| \sum_{ph} (X_{ph}^{l*} + (-1)^l Y_{ph}^{l*})\langle p \| o_l \| h\rangle |^2,
\end{eqnarray}
where $\langle p \| o_l \| h\rangle$ is the reduced matrix element of the single
transition operator, e.g., $o_l=r^3Y_{1\mu}(\hat{\bm{r}})$ for an isoscalar
dipole mode. Because we are discussing the transition between the $0^+$ ground
state and excited state $\nu$ with an angular momentum $J$, we have $l=J$.

The response strength distribution against excitation energy $E$ is given by
\begin{eqnarray}
R(E)=\sum\limits_\nu B(EJ) \delta(E-\hbar\Omega_\nu).
\end{eqnarray}
The above discrete distribution is smoothed using the Lorentzian function to simulate escaping and spreading widths. Finally, we obtain the continuous strength function,
\begin{eqnarray}
  \label{eq:response}
R(E)=\sum\limits_\nu B(EJ) \frac{1}{\pi} \frac{\Gamma/2}{(E-\hbar\Omega_\nu)^2+(\Gamma/2)^2},
\end{eqnarray}
where the width of $\Gamma=2$ MeV is used in this paper. This is a
typical value for $\Gamma$ that has been adopted in the literature~(see e.g.
Ref.~\cite{Paar2006,Papakonstantinou2007}).

The transition density can give detailed information about the resonance, which is defined as the transition amplitude of the density operator. For the state $|\nu\rangle$, the radial transition density is defined as
\begin{eqnarray}
  \delta\rho_\nu^J(r)&=&\sum\limits_\mu \langle \nu | \sum\limits_i \frac{\delta(r-r_i)}{r^2} Y_{J\mu}(\hat{\bm r}_i) |0 \rangle\nonumber\\
  &\approx&\sum_{ph} (X_{ph}^{J*} + (-1)^J Y_{ph}^{J*})\langle p \| \frac{\delta(r-r_i)}{r^2} Y_{J\mu}(\hat{\bm r}_i) \| h\rangle,
\end{eqnarray}
where for the proton (neutron) transition density, the summation runs over the protons (neutrons) only.

\section{Calculations and discussions}

 In the present paper, we use the chiral NNLO$_{\text{sat}}$ potential to
 calculate collective multipole resonances within the framework of the random
 phase approximation. We focus on monopole, dipole and quadrupole resonances
 which have been observed in various experiments. The NNLO$_{\text{sat}}$
 interaction contains the NNN force in the normal-ordered two-body
 approximation~\cite{PhysRevLett.109.052501,PhysRevC.76.034302}. 
 The NNN effects on the giant resonances of light nuclei have been investigated
 with few-body {\it ab-initio} approaches
 \cite{EFROS2000223,PhysRevLett.96.112301,PhysRevC.91.024303, QUAGLIONI2007370}.
 In Refs. \cite{EFROS2000223,PhysRevLett.96.112301}, with the Argonne NN and
 Urbana NNN forces, the authors used the correlated hyperspherical harmonics
 method to calculate the photodisintegrations of the giant dipole resonances in
 $^3$H, $^3$He and $^4$He. The chiral potential with N$^3$LO (NN) and N$^2$LO (NNN) was used for
 the $^4$He giant monopole resonance within the {\it ab-initio} few-body Jacobi
 coordinates \cite{PhysRevC.91.024303} and for the $^4$He giant dipole resonance
 with no core shell model \cite{QUAGLIONI2007370}. The calculations indicate
 that the NNN effects appear mainly for high excitation energies ($E_x>50$ MeV), increasing total resonance cross sections.
\begin{figure}[!htbp]
\centering
\includegraphics[width=\columnwidth]{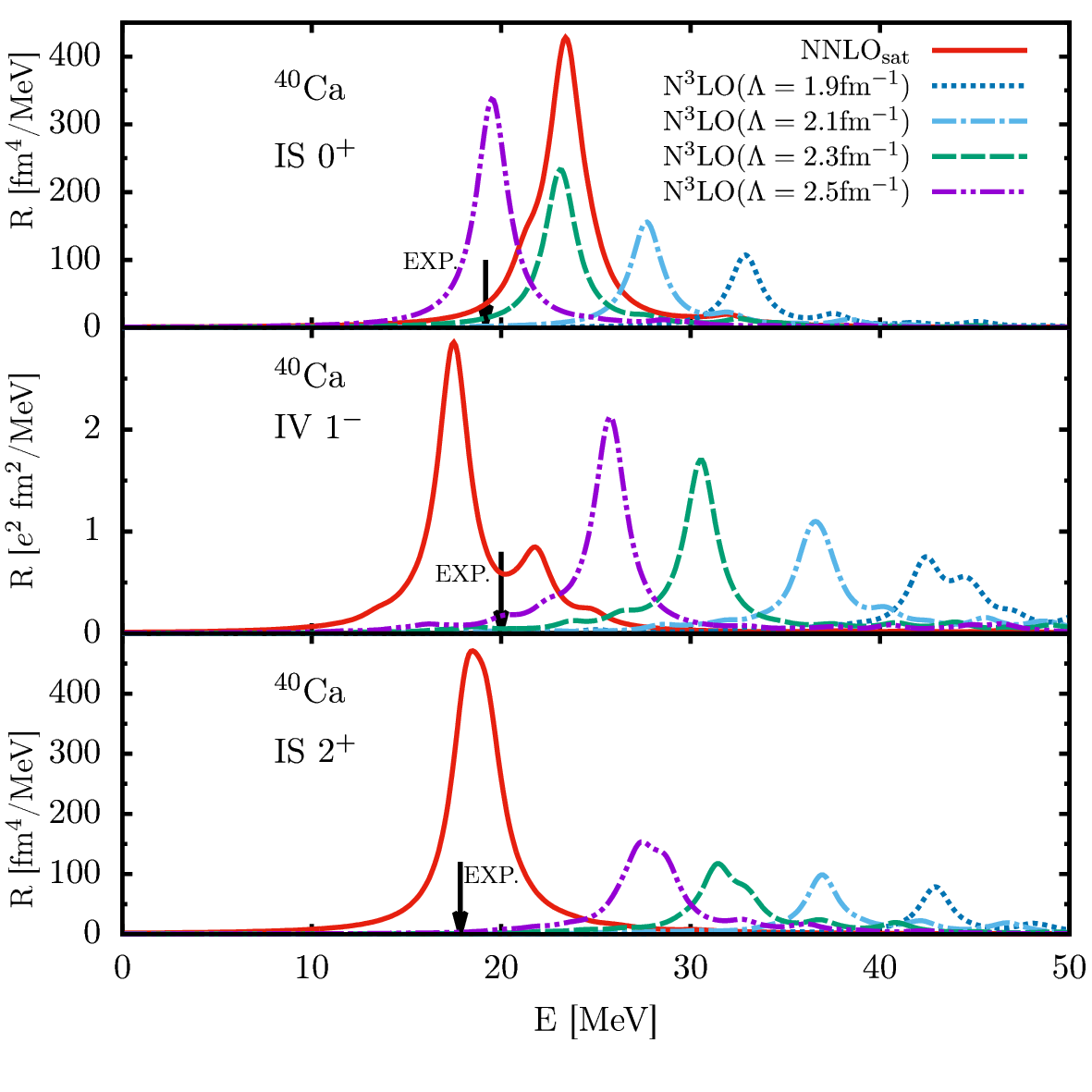}
\caption{\label{fig:Ca40}Calculated $^{40}$Ca isoscalar (IS) and isovector (IV) strength distributions using NNLO$_{\text{sat}}$ and N$^3$LO (at different V$_{\text{low-}k}$ cutoff $\Lambda$). The experimental centroid energies from \cite{PhysRevC.64.049901,Berman1975} are indicated by arrows. The harmonic oscillator basis parameter $\hbar\omega$ and the basis truncation $N_{\text{max}}$ are same as in Fig. \ref{fig:binding}.}
\end{figure}
 
 Fig. \ref{fig:Ca40} shows the NNLO$_{\text{sat}}$-RPA calculations of strength
 distributions for isoscalar giant monopole resonances (ISGMR), isovector giant
 dipole resonances (IVGDR) and isoscalar giant quadrupole resonances (ISGQR) in
 $^{40}$Ca. For comparison, we have also shown results using the chiral N$^3$LO (NN) potential from
 \textcite{Entem2003}, softened to various cutoffs using the $V_{\text{low-}k}$
 method \cite{Bogner2003}.
 We first performed a spherical HF calculation in a harmonic oscillator basis,
 then used the normal-ordered two-body approximation \cite{PhysRevLett.109.052501,PhysRevC.76.034302} to arrive
 at a two-body interaction used in the residual RPA $ph$ calculations.
 We see that the NNLO$_{\text {sat}}$ calculations are in overall agreement with experimental resonance peaks, while the N$^3$LO (NN)
 calculations are sensitive to the $V_{\text{low-}k}$ momentum cutoff parameter
 $\Lambda$. No single $\Lambda$ value is found to describe ISGMR, IVGDR and
 ISGQR simultaneously. The dependence on the softening parameter has been
 commented in the RPA calculations based on the AV18 potential by the unitary correlation
 operator method (UCOM) \cite{Paar2006}. This is a clear indication that induced
 three-body forces must be included for this approach to be valid. With the coupled-cluster method, the bare N$^3$LO (NN)
 interaction was successfully applied to $^4$He, $^{16,22}$O and $^{40}$Ca giant dipole
 resonances \cite{PhysRevLett.111.122502,PhysRevC.90.064619}. Using
 bare forces, there are no induced three-body forces, however explicit
 three-body forces must still be considered.
 
According to the macroscopic interpretation of the giant dipole resonance by the
Goldhaber-Teller \cite{PhysRev.74.1046} or Steinwedel-Jensen model
\cite{PhysRev.79.1019}, the resonance energy is inversely proportional to the
radius of the nucleus. It has been known that {\it ab-initio} calculations based
on realistic two-body interactions neglecting three-body forces give smaller
radii compared with experimental data
\cite{BINDER2014119,PhysRevC.73.044312,PhysRevC.91.051301,Hu2016}. NNLO$_{\text {sat}}$ is
optimized with nuclear radii, which improves the descriptions of nuclear bulk properties including nuclear
electric dipole polarizabilities \cite{Hagen:2016aa,PhysRevC.94.034317}. 

\subsubsection{Monopole resonances}
\begin{figure}[!htbp]
\centering
\includegraphics[width=\columnwidth]{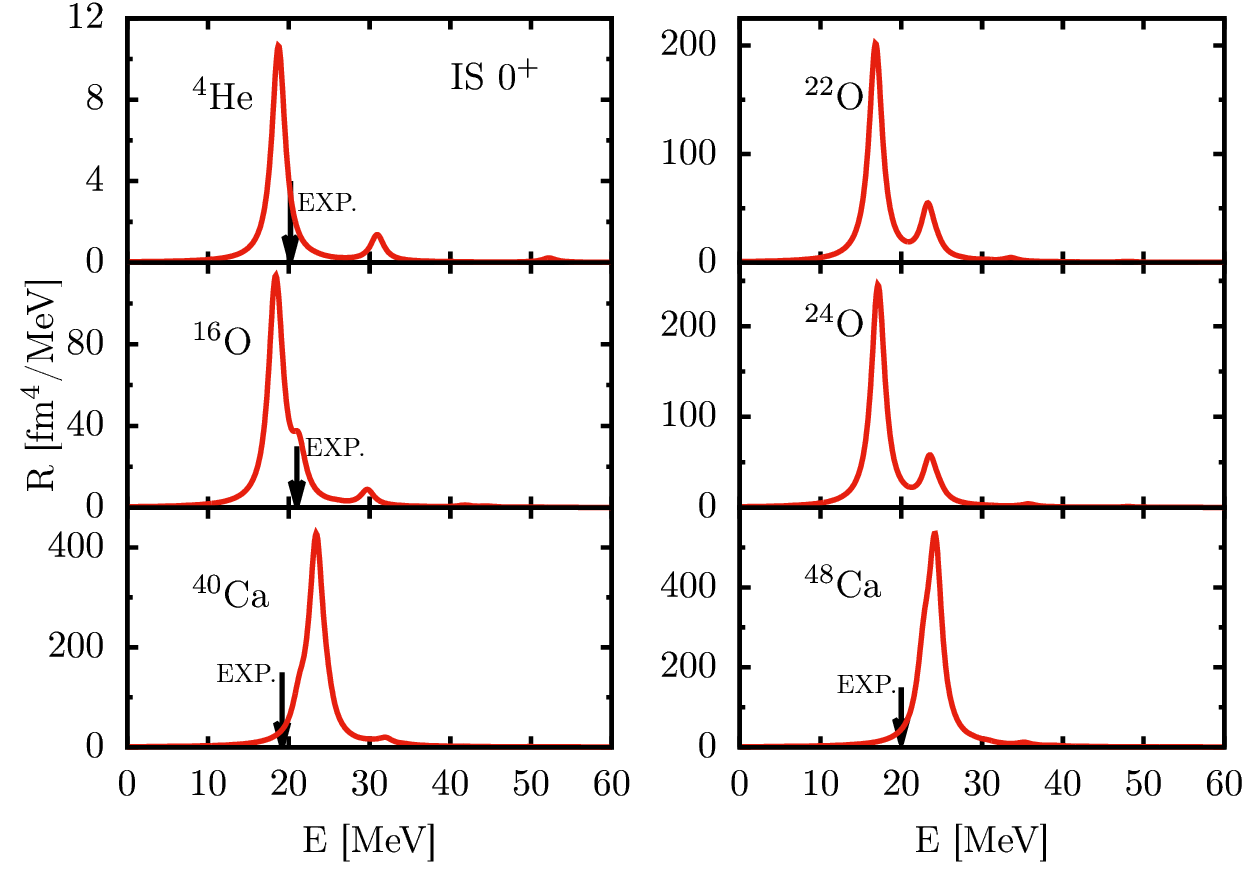}
\caption{\label{fig:monopole}Calculated strength distributions of isoscalar monopole resonance using $\mathrm{NNLO_{sat}}$. The experimental centroid energies are indicated by arrows. Experimental data are taken from Refs.  \cite{PhysRevC.64.064308} ($^{16}$O),  \cite{PhysRevC.64.049901} ($^{40}$Ca),  \cite{PhysRevC.83.044327} ($^{48}$Ca). For $^4$He, the first excited state energy is adopted as the centroid energy \cite{PhysRevC.91.024303}. The harmonic basis parameter and basis truncation are same as Fig. \ref{fig:binding}.}
\end{figure}

The ISGMR is a breathing mode in which neutrons and protons move in phase. The ISGMR excitation energy is connected to the incompressibility of nuclear matter \cite{TREINER1981253}. It provides important information on nuclear matter which cannot be probed directly in ordinary laboratories.

The calculated ISGMR strength distributions using NNLO$_{\text{sat}}$ are displayed in Fig. \ref{fig:monopole} for the closed-shell nuclei, $^4$He, $^{16,22,24}$O and $^{40,48}$Ca. The centroid energies obtained from experimental data with inelastic $(\alpha,\alpha')$ scattering are given for comparison.
For $^4$He, our calculations show that the first $0^+$ excited state has a
breathing excitation mode, consistent with Ref.~\cite{PhysRevC.91.024303}. We see that the present calculations are in reasonable agreement with the available data. The small discrepancies ($\approx$ 1--4 MeV) between the calculations and data could originate from missing correlations in RPA.

\subsubsection{Dipole resonances}

The IVGDR has been the subject of a large number of studies. Many of these
have focused on low-lying dipole strengths in nuclei away from the
valley of stability \cite{0034-4885-70-5-R02,1402-4896-2013-T152-014012}.
Isovector resonances are connected to the symmetry energy and the slope of
the symmetry-energy curve, and thus provide critical information of constraints
on the equation of state of nuclear matter \cite{refId0}.

\begin{figure}[!htbp]
\centering
\includegraphics[width=\columnwidth]{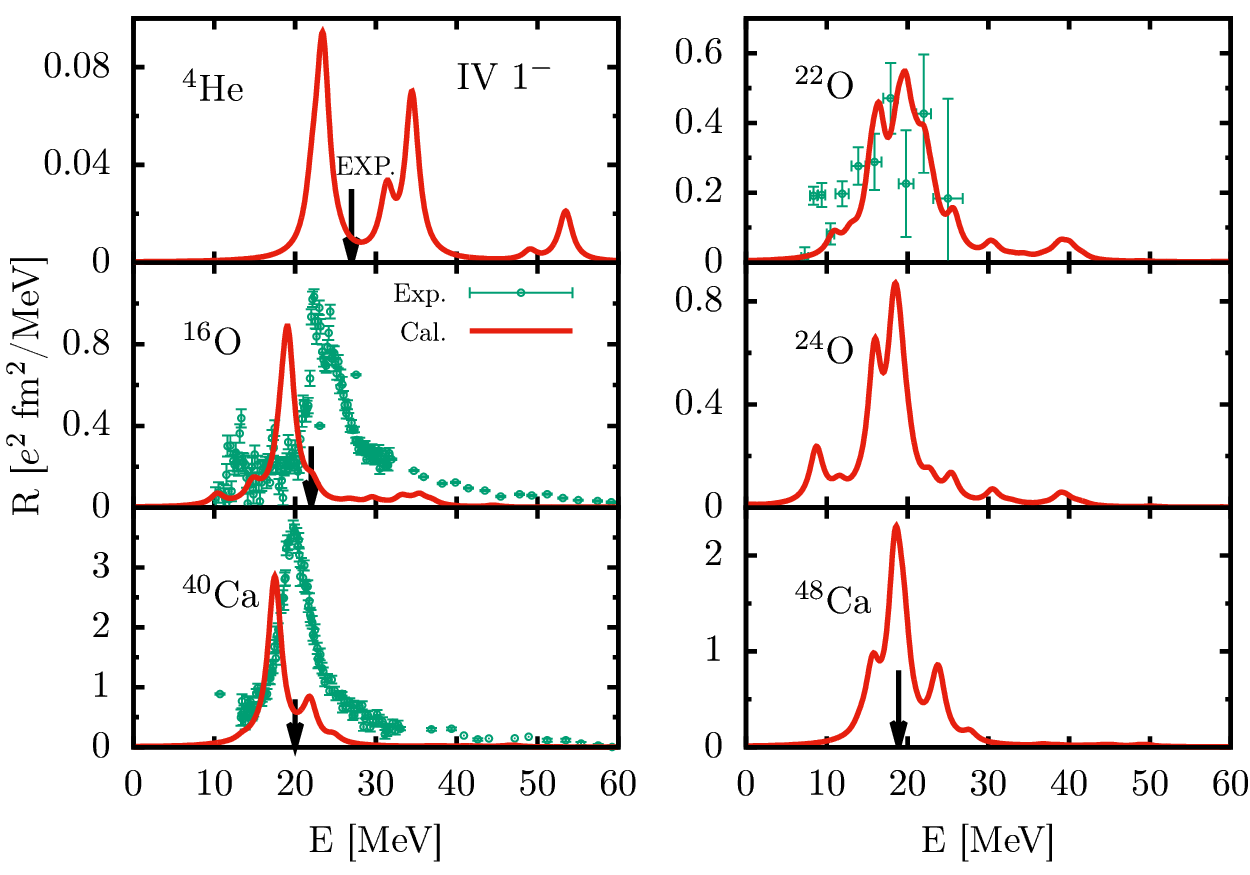}
\caption{\label{fig:dipole}Calculated strength distributions of isovector dipole resonances using NNLO$_{sat}$. The experimental centroid energies are indicated by arrows. Experimental centroids are taken from \cite{Ca48_dipole_exp} for $^{48}$Ca, and \cite{Berman1975} for other nuclei. The experimental photoabsorption cross sections for $^{16}$O, $^{22}$O and $^{40}$Ca are from Refs. \cite{AHRENS1975479,cdfe,PhysRevLett.86.5442}. The harmonic oscillator basis parameter $\hbar\omega$ and the basis truncation $N_{\text{max}}$ are same as in Fig. \ref{fig:binding}.}
\end{figure}

In Fig. \ref{fig:dipole}, the calculated isovector dipole strength distributions
are displayed and compared to experimental data. The calculations are in
reasonable agreement with experiments. For $^{16,22}$O and $^{40}$Ca, the
strength functions extracted from experimental photoabsorption cross sections
are also displayed. The centroid energies obtained from RPA calculations are about
1--3 MeV lower than the experimental data. This may be improved by including
additional many-body correlations.

The plots for $^{22}$O and $^{24}$O show evidence of low-lying strengths around
10 MeV. It is more pronounced in $^{24}$O. This low-energy resonance is usually
called the pygmy dipole resonance (PDR), which is interpreted as the oscillation
of excess neutrons against the $N=Z$ core. In our calculations, the strength
below 15 MeV exhausts about 4.3\% of the classical Thomas-Reiche-Kuhn (TRK) sum
rule in $^{22}$O, while 10.7\% in $^{24}$O. The experimental data of $^{22}$O
give about 8\% of the classical TRK sum rule up to excitation energy of 15 MeV
\cite{PhysRevLett.86.5442}.

\begin{figure}[!htbp]
\centering
\includegraphics[width=\columnwidth]{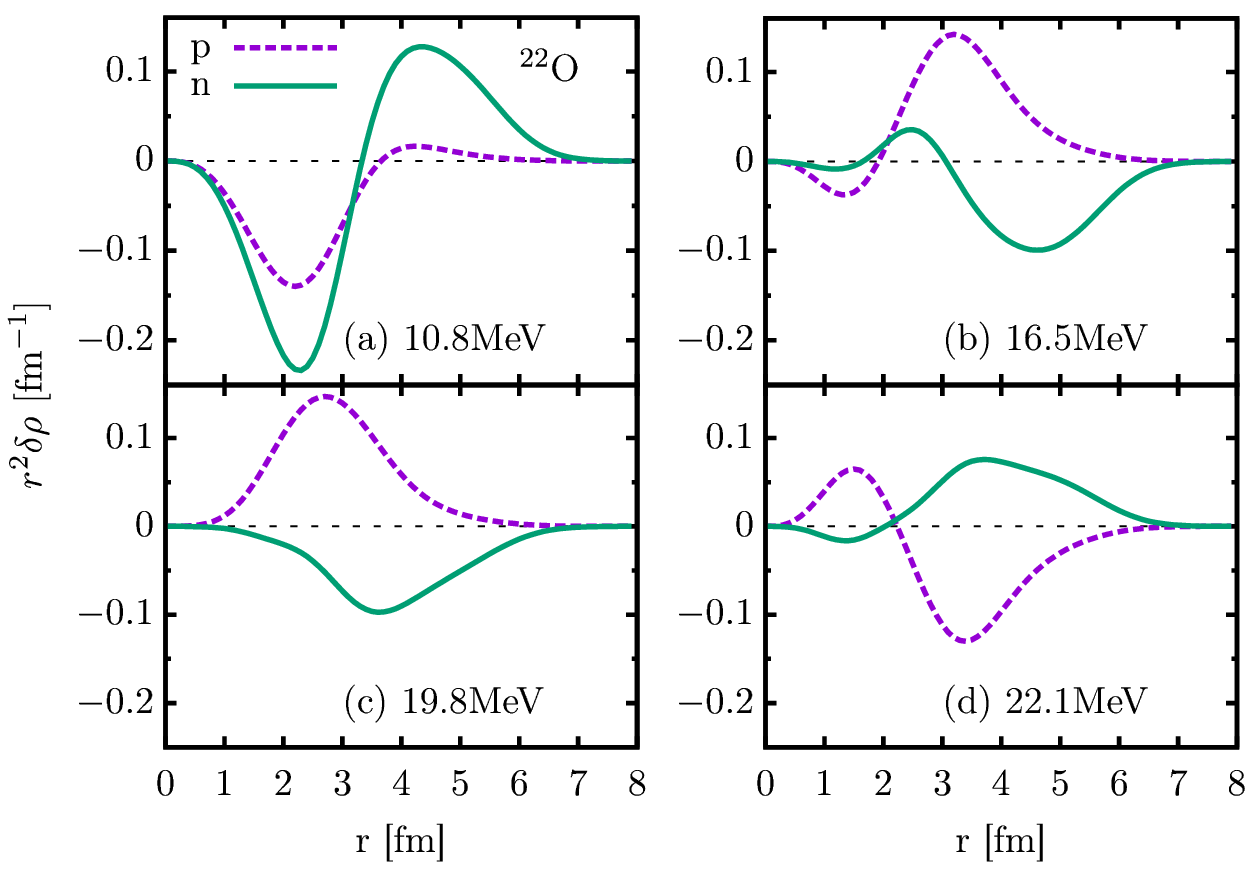}
\caption{\label{fig:tr_density_O22}Calculated $^{22}$O transition densities for the protons and neutrons at the prominent peaks of the dipole resonance: (a) 10.8 MeV, (b) 16.5 MeV, (c) 19.8 MeV and (d) 22.1 MeV.}
\end{figure}

\begin{figure}[!htbp]
\centering
\includegraphics[width=\columnwidth]{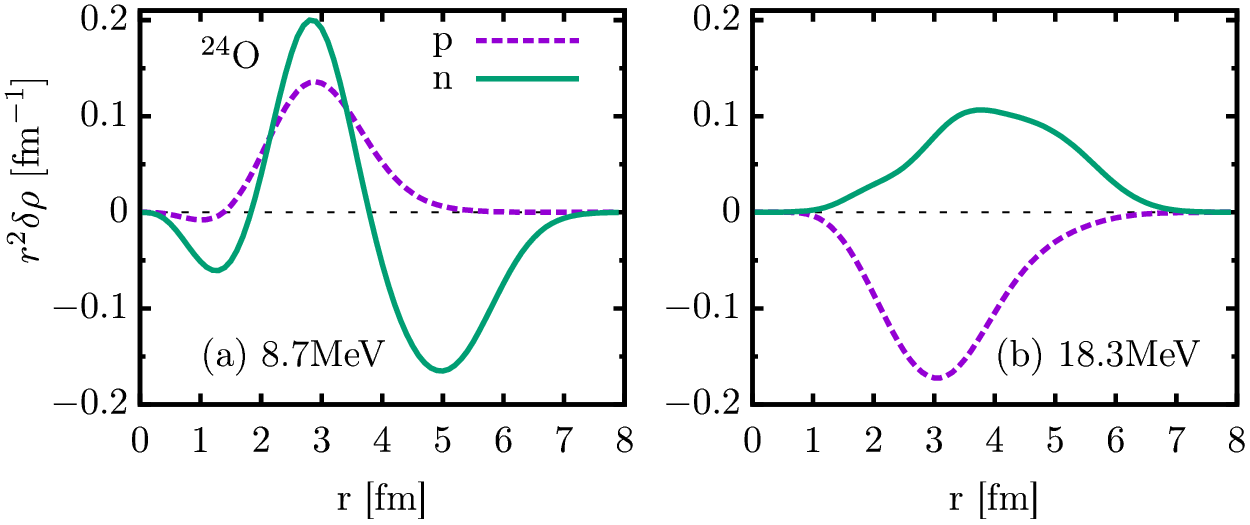}
\caption{\label{fig:tr_density_O24}Calculated $^{24}$O transition densities for the protons and neutrons at the most prominent peaks of the dipole resonance: (a) 8.7 MeV and (b) 18.3 MeV.}
\end{figure}

\begin{figure}[!htbp]
\centering
\includegraphics[width=0.8\columnwidth]{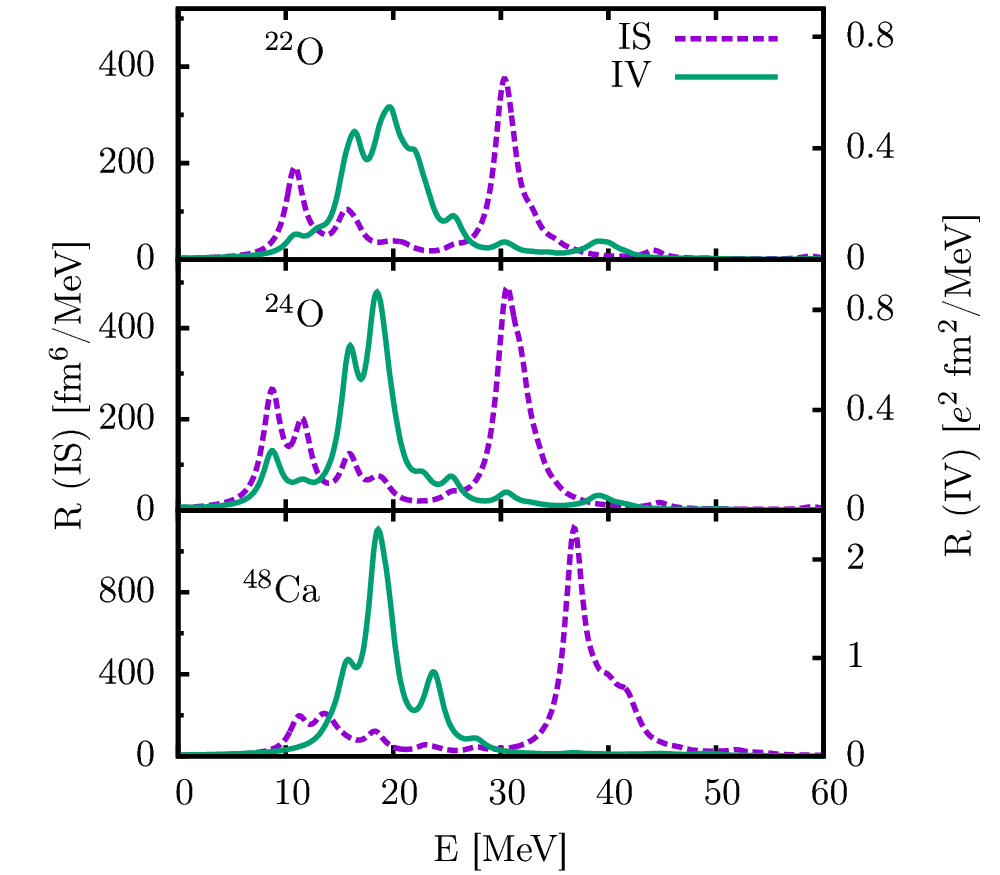}
\caption{\label{fig:ISD}Comparison of isoscalar and isovector dipole resonance strength distributions. For visibility, different scales have been used for the isoscalar (left) and isovector (right) channels.}
\end{figure}

To investigate the nature of these resonances, we calculate transition densities
shown in Figs. \ref{fig:tr_density_O22} and \ref{fig:tr_density_O24} for
$^{22}$O and $^{24}$O at the prominent peaks. For the $^{22}$O peak at 10.8 MeV,
the protons and neutrons oscillate in phase in the interior, while in the
surface region the neutron transition density dominates. This is consistent with
a PDR. At the higher resonance peaks, the transition densities show that the
protons and neutrons oscillate out of phase, exhibiting the typical and
well-known picture of a GDR. Fig. \ref{fig:tr_density_O24} shows a similar
behavior in the transition densities of $^{24}$O. The present results are
consistent with the relativistic RPA calculations given in Refs.
\cite{VRETENAR2001496,PhysRevC.67.034312}. There, it was discussed that
low-energy soft dipole resonances in halo nuclei have similar in-phase
transition densities \cite{halo_multipole,PhysRevC.89.064303}. The soft dipole
resonance appears due to the oscillation of the halo neutrons against a core.
The excitation energy of the soft resonance is low, e.g., $\approx 2-3$ MeV in
$^6$He, $^{11}$Li and $^{12}$Be \cite{halo_multipole,PhysRevC.89.064303}. The
low-energy dipole strength in light neutron-rich nuclei is interpreted as mainly
arising from single-neutron transitions rather than showing much collectivity
\cite{1402-4896-2013-T152-014012,Sagawa1995}. The oxygen isotopes $^{22, 24}$O
have larger neutron separation energies than the loosely bound light nuclei,
thus the low-energy strengths are less pronounced because of the threshold effect \cite{CATARA1996181}.

The RPA formalism also provides a way to analyze the wave function of a
peak state. For $^{22}$O, the main components of the wave function for the
excited state at 10.8 MeV are the neutron excitations: 62\% $0d_{5/2}\rightarrow
1p_{3/2}$, 15\% $0p_{1/2}\rightarrow 1s_{1/2}$ and 8\% $0d_{5/2}\rightarrow
0f_{7/2}$, while relativistic RPA calculations give 93\% $0d_{5/2}\rightarrow
1p_{3/2}$ and 3\% $0d_{5/2} \rightarrow 0f_{7/2}$ \cite{VRETENAR2001496}. For
$^{24}$O, the main components for the state at 8.7 MeV are the neutron
excitations: 87\% $1s_{1/2}\rightarrow 1p_{3/2}$ and 5\% $0d_{5/2}\rightarrow
0f_{7/2}$, while relativistic RPA calculations give 93\% for the neutron
particle-hole excitation $1s_{1/2}\rightarrow 1p_{3/2}$.

The PDR mode is of special interest in neutron-rich nuclei
\cite{PhysRevC.96.031301}. The transition densities indicate that the low-lying
strengths are mainly of isoscalar nature, see the subfigures (a) in Figs.
\ref{fig:tr_density_O22} and \ref{fig:tr_density_O24}. Fig. \ref{fig:ISD},
however, contrasts the isoscalar dipole resonance (isoscalar compressional dipole
mode) strength distributions with the isovector dipole resonances for $^{22}$O,
$^{24}$O and $^{48}$Ca, and shows that for  $^{22}$O and $^{24}$O, the low-energy
peaks of the isoscalar and isovector channels appear at the same energy. This
indicates that the PDR excitation has a nature of isoscalar and isovector
mixing. The GDR peaks at higher energy does not exhibit the same mixing,
although the transition densities at the isoscalar GDR peaks look similar to 
those of the proposed PDR peaks in $^{22}$O and $^{24}$O.
The strength distributions for $^{48}$Ca does not display the same mixing, as
only an isoscalar peak is found at low-energy around 11 MeV.
Further experimental studies are necessary to verify the low-energy mixing in
$^{22}$O and $^{24}$O and should be obtainable in inelastic $\alpha$ scattering
experiments.

\subsubsection{Quadrupole resonances}
\begin{figure}[!htbp]
\centering
\includegraphics[width=\columnwidth]{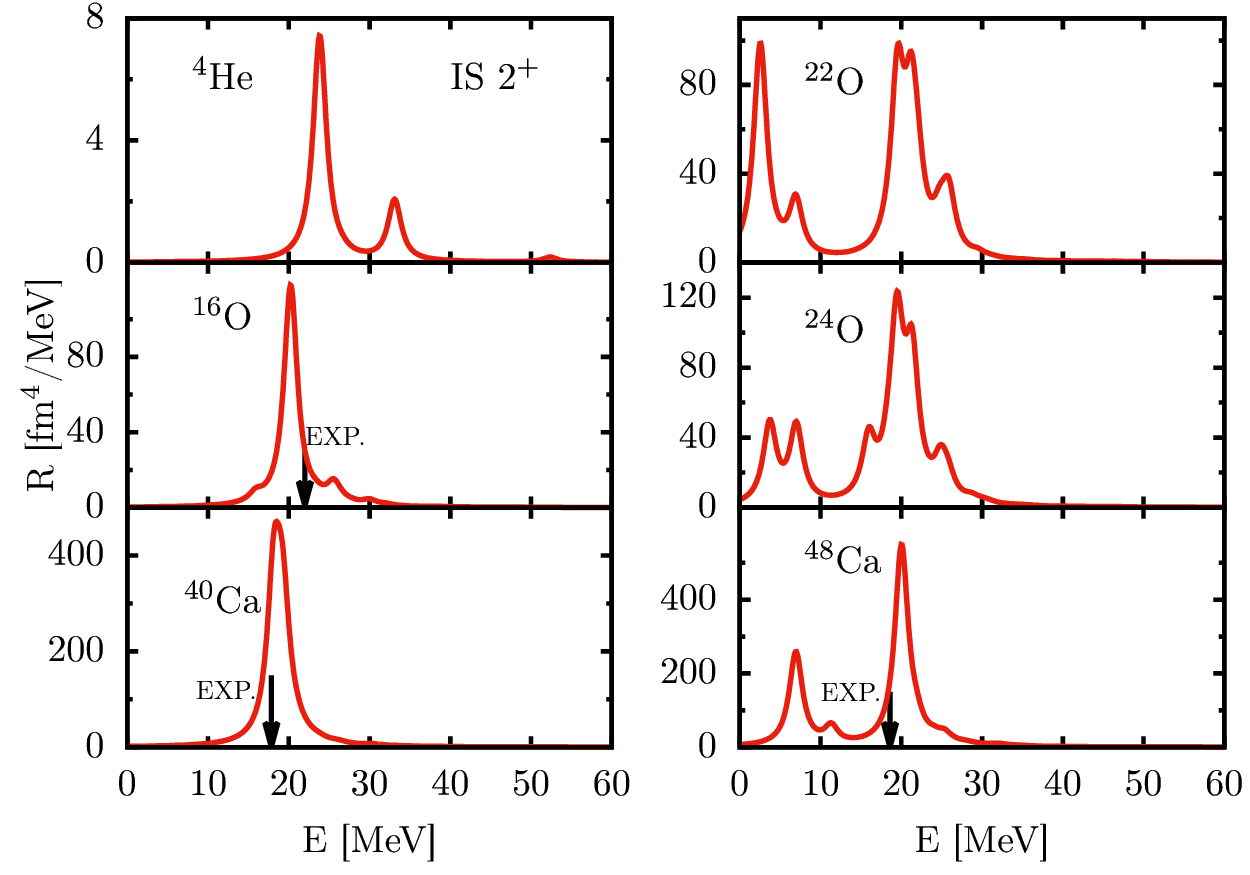}
\caption{\label{fig:quadrupole}Calculated strength distributions of isoscalar quadrupole resonance using $\mathrm{NNLO_{sat}}$. The experimental centroid energies are indicated by arrows. Experimental data are taken from Refs.  \cite{PhysRevC.64.064308} ($^{16}$O),  \cite{PhysRevC.64.049901} ($^{40}$Ca),  \cite{PhysRevC.83.044327} ($^{48}$Ca). The harmonic basis parameter and basis truncation are same as in Fig. \ref{fig:binding}.}
\end{figure}

Fig. 8 shows the calculated isoscalar quadrupole strength distributions, as well
as the available experimental centroid energies. The calculations are in good
agreement with experimental data where available. It is interesting that the neutron-rich $^{22,24}$O and
$^{48}$Ca have pronounced low-energy peaks in the calculated quadrupole
strengths. The low-lying peak should be compared with the experimental
excitation energy of the corresponding $2^+$ state. We have analyzed the RPA
wave function, and find that the low-lying peak is predominantly a
single-neutron excitation. The calculated peak at 2.5 MeV in $^{22}$O is
a one-neutron excitation from $0d_{5/2}$ to $1s_{1/2}$, compared with the
excitation energy of the observed first $2^+$ state at 3.2 MeV \cite{O22_exp}.
For $^{24}$O, the present calculations give that the low-energy peaks at 3.7 and
7.0 MeV are due to the single-neutron excitations from $1s_{1/2}$ to $0d_{3/2}$
and from $0d_{5/2}$ to $0d_{3/2}$, respectively, while the experimental $2^+_1$
state energy is 4.7 MeV \cite{O24_exp}.

The present NNLO$_{\text{sat}}$-RPA calculated excitation energies of the
low-lying $2^+$ resonance states are also consistent with our previous
calculations using the Gamow shell model (GSM) with the CD-Bonn potential
\cite{SUN2017227} . The Gamow shell model takes into account resonance and
continuum by using a complex-momentum Berggren coordinates. The GSM calculations
give that $^{22}$O has a $2^+$ excited state at 3.3 MeV with a one-neutron
excitation from $0d_{5/2}$ to $1s_{1/2}$ as the dominant configuration. The
nucleus $^{24}$O has a $2^+$ resonance state at 4.5 MeV, dominated by
a one-neutron excitation from $1s_{1/2}$ to $0d_{3/2}$ \cite{SUN2017227} . The
energy gap between neutron $1s_{1/2}$ and $0d_{3/2}$ is larger than the gap
between neutron $0d_{5/2}$ and $1s_{1/2}$. A decrease of the $B(E2)$ values from
$^{22}$O to $^{24}$O is obtained in our RPA calculations, consistent with a $1s_{1/2}$
subshell closure ($N=16$).

\section{Summary}
In the present work, the chiral NNLO$_{\text{sat}}$ potential, which includes a
three-body force, has been used to investigate the monopole, dipole, and
quadrupole resonances of the closed-shell nuclei $^4$He, $^{16, 22, 24}$O and
$^{40, 48}$Ca. The calculations were done in the HF-RPA approach. Due to strong
short-range correlations, the HF calculation based on realistic nuclear forces
cannot give right binding energies of nuclei. The energy of the ground state is
calculated by the RPA energy and a correction from second-order perturbation
theory. The RPA NNLO$_{\text{sat}}$ calculations reproduce the available
experimental centroid energies of isoscalar monopole, isovector dipole and
isoscalar quadrupole resonances of these nuclei reasonably well. The HF-RPA
resonance calculations using a softened two-body realistic interaction without
three-body force are sensitive to the softening parameter, indicating the
importance of induced three-body force.

In neutron-rich $^{22}$O and $^{24}$O, we obtain low-energy dipole resonances at
excitation energies around 10 MeV. The calculations of response strengths show
that the low-energy resonances are a mix between isoscalar and
isovector dipole resonances. However, the transition density calculations
indicate that the low-energy dipole resonances are dominated by the isoscalar mode, while
the dipole resonances at higher energies exhibit the characteristic of the
isovector mode. The RPA calculations reveal that the low-energy dipole resonance
is dominated by a single-neutron excitation, which is consistent with other RPA
calculations based on phenomenological interactions. Prominent peaks at low
energies between 2 and 7 MeV are found in the isoscalar quadrupole response
functions of neutron-rich $^{22, 24}$O and $^{48}$Ca. The low-energy quadrupole
resonances are identified as single-neutron transitions. The peak positions are
in reasonable agreement with the experimental energies of the corresponding
$2^+$ states and Gamow shell model calculations. The RPA calculation may be
improved by including high-order correlations ( e.g. $2p2h$, $3p3h\cdots$ excitations).

\begin{acknowledgments}
  Valuable discussions with T. Papenbrock, U. Garg, J. P. Vary and J.C. Pei are gratefully
  acknowledged. This work has been supported by the National Natural Science
  Foundation of China under Grants No. 11235001, No. 11320101004 and No.
  11575007; and the CUSTIPEN (China-U.S. Theory Institute for Physics with
  Exotic Nuclei) funded by the U.S. Department of Energy, Office of Science
  under Grant No. DE-SC0009971.  This work was partially supported by the Office
  of Nuclear Physics, U.S. Department of Energy, under Grants DE-FG02-96ER40963,
  DE-SC0008499 (NUCLEI SciDAC collaboration), the Field Work Proposal ERKBP57 at
  Oak Ridge National Laboratory (ORNL), and used resources of the Oak Ridge
  Leadership Computing Facility located at ORNL, which is supported by the
  Office of Science of the Department of Energy under Contract No.
  DE-AC05-00OR22725. Computer time was partially provided by the Innovative and Novel
  Computational Impact on Theory and Experiment (INCITE) program.
\end{acknowledgments}

\bibliography{rpa}

\end{document}